\documentclass[twocolumn]{aastex701}

\newcommand{\h}{$h_{70}^{-1}$}

\newcommand{\be}{\begin{equation}}
\newcommand{\ee}{\end{equation}}
\newcommand{\ba}{\begin{eqnarray}}
\newcommand{\ea}{\end{eqnarray}}

\newcommand{\simgt}{\lower 2pt \hbox{$\, \buildrel {\scriptstyle >}\over {\scriptstyle\sim}\,$}}
\newcommand{\simlt}{\lower 2pt \hbox{$\, \buildrel {\scriptstyle <}\over {\scriptstyle\sim}\,$}}
\newcommand{\ls}{\lower 2pt \hbox{$\;\scriptscriptstyle \buildrel<\over\sim\;$}}
\newcommand{\gs}{\lower 2pt \hbox{$\;\scriptscriptstyle \buildrel>\over\sim\;$}}

\begin{document}

\title{DESI Interacting Dwarfs Survey: Multiple Evolutionary Channels of Dwarf Galaxies Driven by Major Dwarf Interactions}

\author[]{Marko Mi\'ci\'c}
\affil{Homer L.\ Dodge Department of Physics and Astronomy,
University of Oklahoma, Norman, OK 73019, USA}
\email{micic@ou.edu}
\author[]{Sugata Kaviraj}
\affil{Centre for Astrophysics Research, University of Hertfordshire, College Lane, Hatfield AL10 9AB, UK}
\email{}


\begin{abstract}
Dwarf–dwarf interactions are a major driver of galaxy evolution in the early Universe, but observational constraints on their impact across mass, separation, and environment remain scarce. Using the Dark Energy Spectroscopic Instrument Early Data Release, this study constructs the largest sample of dwarf galaxy pairs, spanning stellar masses $6.5 \leq \log(M_\star/M_\odot) \leq 9.5$ and redshifts $0.015 \leq z \leq 1.5$. We analyze the evolution of star formation rates and multiwavelength colors ($NUV–r$, $U$–$V$, $V$–$J$, $r$–$K$, and $g$–$r$) as a function of projected separation, stellar mass, and external tidal environment. Low-mass/low-redshift pairs exhibit interaction-triggered star formation at intermediate separations. At the smallest separations ($\lesssim$10 kpc), these systems display divergent behavior: isolated pairs show the onset of dust-obscured star formation, those embedded in weak external tidal fields undergo extended, dust-unobscured star formation, while those embedded in strong external tidal fields experience rapid quenching of star formation. High-mass/low-redshift pairs follow a more uniform evolutionary track, with generally weaker dependence on environment. Interaction-triggered star formation occurs at intermediate separations, followed by increased dust obscuration and rapid quenching at the smallest separations. At higher redshifts, high-mass dwarf systems display an initial suppression of star formation, followed by renewed activity at smaller separations, consistent with early-stage gas removal and later interaction-driven inflows. These results demonstrate that dwarf–dwarf interactions do not follow a single evolutionary pathway. Instead, we reveal a complex picture of dwarf galaxy evolution, with significant implications for dust-obscured star formation, abrupt quenching, and the formation of compact, dusty, intensely star-forming systems.
\end{abstract}


\section{Introduction} 
Dwarf galaxies ($\log(M_\star/M_\odot)\leq9.5$) are the most abundant type of galaxies in the Universe and are widely regarded as the fundamental building blocks of galaxy formation \citep[e.g.,][]{1990A&A...228...42B,1994A&ARv...6...67F,1998ARA&A..36..435M}. They are expected to be particularly prevalent in the early Universe, where they dominate the merger rate per unit volume, undergoing multiple concurrent and consecutive interactions with other dwarfs \citep{ 2015ApJ...813...21M,2015ApJ...800...12C,2024Natur.626..975A}. Therefore, low-mass galaxy interactions and mergers play a key role in the early stages of hierarchical galaxy assembly, providing a direct insight into how present-day galaxies formed. Gravitational interactions between dwarf galaxies can induce bursts of star formation, redistribute gas and dark matter, transform morphologies, and accelerate chemical enrichment \citep[e.g,][]{2015ApJ...805....2S}. Additionally, dwarf–dwarf interactions provide a unique laboratory for studying the interplay between gravity, gas physics, and star formation \citep{2016MNRAS.459.1827P,2021MNRAS.502.4262J}.

Star formation in dwarf galaxies is particularly sensitive to the availability and state of their gas. Their shallow gravitational potentials make their gas reservoirs strongly susceptible to stellar feedback, resulting in bursty and highly variable star formation histories \citep[e.g.,][]{2007ApJ...667..170S,2014MNRAS.445..581H,2016ApJ...820..131E}. In interacting systems, tidal forces provide an external mechanism to perturb gas reservoirs and drive inflows, offering a clear window into how the interstellar medium responds to gravitational encounters \citep[e.g.,][]{2017ApJ...846...74P}. Furthermore, low-mass systems are highly susceptible to external tidal fields and environmental stripping from neighboring massive galaxies \citep[e.g.,][]{2006MNRAS.369.1021M,2016MNRAS.459.1827P}, meaning dwarf pair evolution is governed by both mutual gravitational interactions and broader environmental influence.

Despite their importance, dwarf–dwarf interactions remain poorly explored from an observational point of view. Most past studies of galaxy interactions have focused on massive galaxies, where tidal features are easier to detect, and large spectroscopic samples have been available \citep[e.g.,][]{2008AJ....135.1877E,2012MNRAS.425.2313T,2019ApJ...881..119P,2022ApJ...940....4S}. Only recently have systematic efforts begun to identify and characterize interacting dwarf systems. Pioneering works by \cite{2015ApJ...805....2S} and \cite{2018ApJS..237...36P} demonstrated that dwarf–dwarf interactions can trigger enhanced star formation. Additionally, \cite{2017NatAs...1E..25S} compiled a small catalog of dwarf galaxy groups, \cite{2020AJ....159..103K}, \cite{2025A&A...702A.222C}, and \cite{2024MNRAS.533.3771L} studied dwarf post-mergers, while \cite{2024ApJ...968L..21M} investigated nuclear activity in paired and grouped dwarfs. However, these studies mostly focus on low-redshift, predominantly high-mass dwarfs and are limited by small sample sizes. As a result, there is a lack of substantial samples spanning both low-mass dwarfs at low redshifts and the general population of dwarfs at earlier cosmic times. This gap significantly limits our ability to trace how the impact of dwarf interactions on galaxy evolution changes with mass and redshift, and to connect present-day dwarf populations with their high-redshift analogs.

In this work, we address these problems by exploiting the Early Data Release (EDR) of the Dark Energy Spectroscopic Instrument (DESI). Using DESI’s spectroscopic depth and sky coverage, we identified over 1,000 dwarf galaxy pairs and groups, increasing the number of known systems by an order of magnitude. This expanded sample size enables a detailed investigation of dwarf–dwarf interactions, providing first insights into the low-mass and high-redshift ends of the dwarf-interacting population.

Throughout the paper, we assume the cosmological parameters of H$_0$ = 70 km s$^{-1}$Mpc$^{-1}$, $\Omega_m$ = 0.3, and $\Omega_\Lambda$ = 0.7. Unless otherwise specified, uncertainties on median quantities are estimated by bootstrap resampling, with the reported uncertainties corresponding to the standard deviation of the resulting distribution of median values.

\section{Sample Selection and Data Analysis}
\begin{figure}[ht!]
\plotone{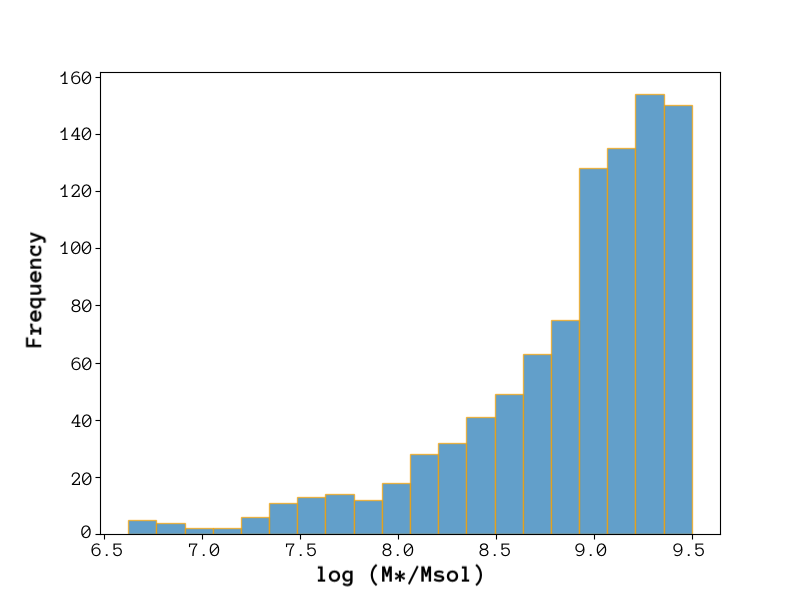}
	\caption{Stellar mass distribution of galaxies involved in major dwarf interactions.
	\label{fig1}}
\end{figure}

\begin{figure}[ht!]
\plotone{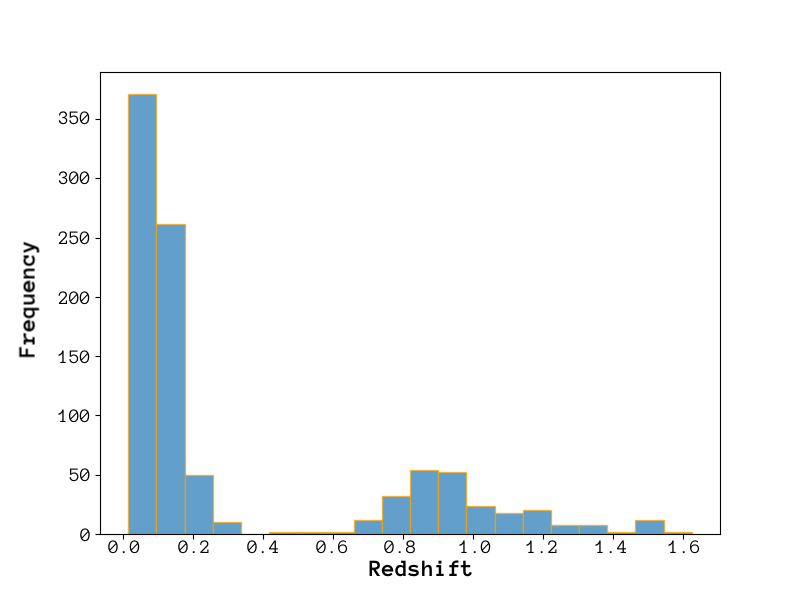}
	\caption{Redshift distribution of galaxies involved in major dwarf interactions. \label{fig2}}
\end{figure}

\begin{figure*}[ht!]
	\plotone{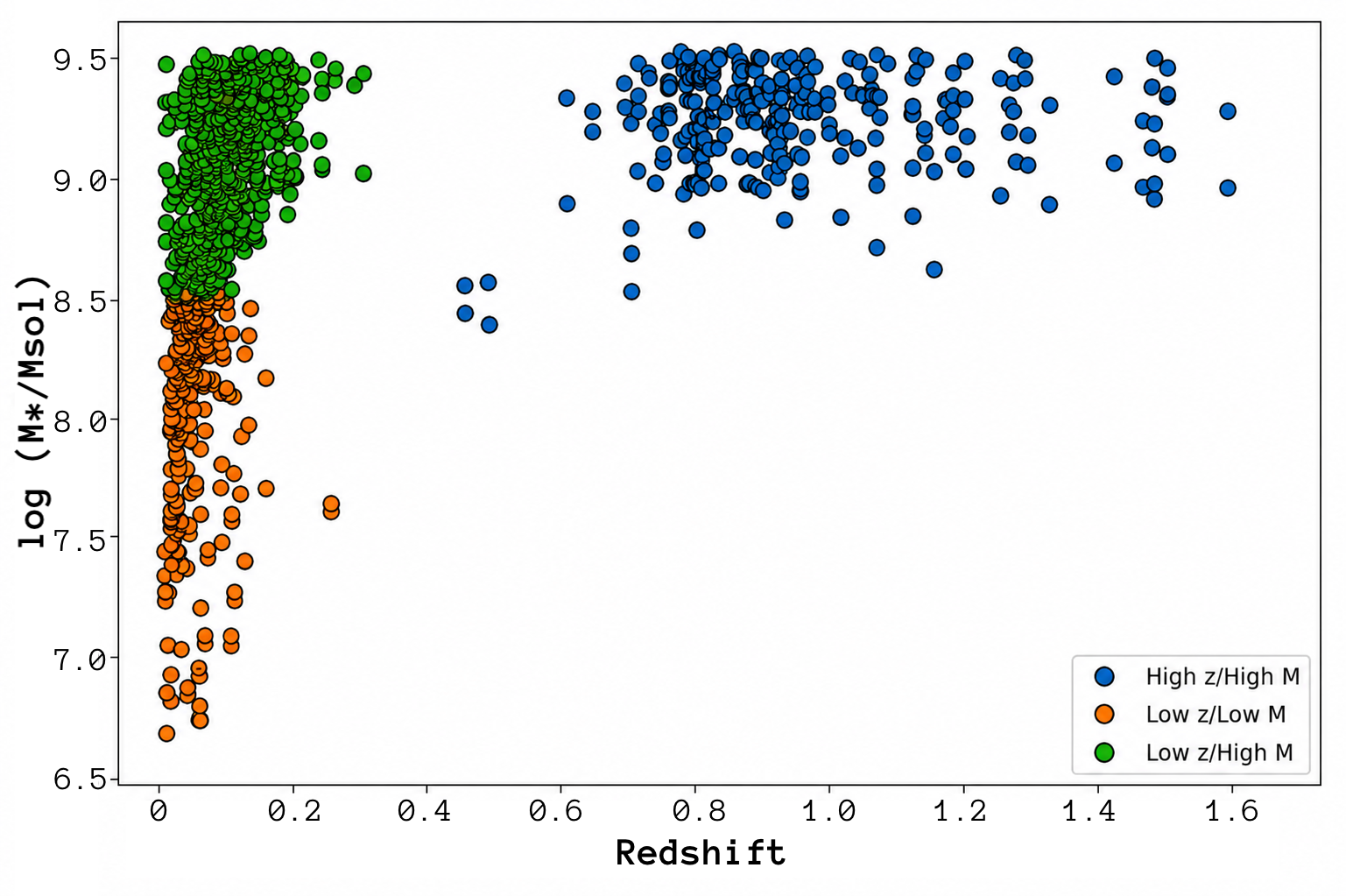}
	\caption{Distribution of galaxies in the stellar mass vs. redshift space. Three distinct subsets are labeled with different colors. \label{fig:spec}}
\end{figure*}
\subsection{DESI Survey}
DESI is a wide-area spectroscopic survey designed to map the large-scale structure of the Universe and measure the expansion history by delivering redshifts of more than 30 million galaxies and quasars  \citep{2016arXiv161100036D}. DESI is installed on the 4-meter Mayall Telescope at Kitt Peak National Observatory and employs a 5,000-fiber robotic positioner system that feeds ten spectrographs, enabling simultaneous observations of thousands of targets over a 3-degree-diameter field of view. The EDR provides the first public spectroscopic dataset from DESI, based on observations obtained during survey validation and the initial stages of the main survey \citep{2024AJ....168...58D}. 

For our analysis, we used the value-added DESI EDR catalog, which provides redshifts, quality flags, star formation rates, stellar masses, and rest-frame photometric measurements for more than a million galaxies \citep{2024A&A...691A.308S}, and can be accessed through the VizieR Catalogue Service \citep{2024yCat..36910308S}. The physical properties were derived using the CIGALE spectral energy distribution (SED) fitting code \citep{2020AAS...23522801B}. The SED modeling combines optical photometry with available WISE mid-infrared photometry and simultaneously accounts for stellar emission, star-formation histories, dust attenuation, and a possible AGN contribution using AGN templates. For more information, we refer the reader to the \cite{2024A&A...691A.308S} paper.
\subsection{Selection Criteria}
We define dwarf galaxies as systems with stellar mass in the range of  $ 6.5\leq\log(M_\star/M_\odot)\leq9.5$. The upper limit is commonly used in the literature as the cut-off mass for dwarf galaxies \citep[e.g.,][]{2024MNRAS.527.1962B,2025ApJ...994....1E}. In contrast, we remove galaxies with extremely low stellar masses near the survey completeness limit to ensure reliable mass estimates across the full sample.

Two dwarfs are classified as a pair if they meet the projected separation, the redshift proximity, and the isolation criterion. Specifically, we require the projected separation to be $r_{proj}<100$ kpc, motivated by the expected tidal influence and virial extents of dwarf galaxy halos \citep{2015ApJ...805....2S}. For stellar masses in the range of $ 6.5\leq\log(M_\star/M_\odot)\leq9.5$, the typical virial radii are of the order of 30-80 kpc. Adopting an extended 100 kpc threshold, therefore, captures both close, strongly interacting systems and wider-separation pairs that are likely on bound orbits or in early interaction stages, and provides a proper baseline for evaluating the impact of interactions as tidal forces start to kick in and become more prominent. Regarding redshift proximity, we require the dwarfs to be within the redshift difference of $\Delta z<(1+z)10^{-4}$, which is the typical redshift uncertainty of the DESI spectroscopic measurements \citep{2016arXiv161100036D,2023AJ....165..144G}. Finally, we require dwarf pairs not to have non-dwarf neighbors within 1 Mpc in projected distance. This ensures that paired dwarfs evolve in isolation, with no external impact from massive galaxies, and that mutual gravitational interactions are the only forces shaping their evolution \citep{2013A&A...560A...9A,2014ApJ...795...54C}.
\subsection{Sample}
We identified 1192 dwarf pairs and 72 dwarf groups consisting of three to seven members. The dwarf pairs and groups span a wide range of stellar masses ($ 6.5\leq\log(M_\star/M_\odot)\leq9.5$), redshifts (0.015 $< z <$ 1.5), and mass ratios (1 $<M_{1*}/M_{2*}<$ 1000). This work focuses on isolated major dwarf–dwarf interactions, in which the galaxies involved have comparable masses ($M_{1*}/M_{2*}<3$), and contains 461 interacting pairs. We will explore more complex interaction scenarios, including minor mergers and dwarf galaxy groups, in follow-up publications.

The stellar mass and redshift distributions of the major-interacting dwarfs are shown in Figure \ref{fig1} and Figure \ref{fig2}, respectively. Since the mass and redshift span a wide range, we further split the galaxies into low-mass/low-redshift ($6.5<\log$ (M$_*$/M$_\odot)<8.5, z<0.4$), high-mass/low-redshift ($8.5<\log$ (M$_*$/M$_\odot)<9.5, z<0.4$), and high-mass/high-redshift ($8.5<\log$ (M$_*$/M$_\odot)<9.5, z>0.4$) bins to test the mass- and redshift-dependence of interaction-induced processes. As evident from Figure \ref{fig2}, the redshift distribution is bimodal, offering a natural border between low- and high-redshift galaxies at $z=0.4$. The bimodality arises from the combination of multiple DESI target classes contributing to the sample, each with different observational selection criteria. Regarding stellar mass, the choice of $ \log(M_\star/M_\odot)=8.5$ as the border value is motivated by the lack of substantial samples of interacting dwarfs with $ \log(M_\star/M_\odot)\leq8.5$ in previous works. For example, \cite{2015ApJ...805....2S} finds only ten such dwarf pairs, which is insufficient to investigate whether low-mass dwarfs behave differently from high-mass dwarfs.

Following the mass ratio, redshift, and stellar mass cuts, our low-mass/low-redshift, high-mass/low-redshift, and high-mass/high-redshift subsets contain 85, 251, and 125 interacting pairs, respectively. Our sample contains no low-mass/high-redshift interacting pairs, as such galaxies become too faint and increasingly difficult to observe and characterize at greater distances. We further divide the three subsets into projected separation bins: 75-100 kpc, 50-75 kpc, 25-50 kpc, 10-25 kpc, and 0-10 kpc. The progressively narrower bins toward smaller separations provide finer sampling of the late interaction stages, where interaction-driven changes are expected to become more pronounced and evolve rapidly \citep{2011MNRAS.412..591P,10.1111/j.1365-2966.2012.21749.x}. The distribution of dwarfs from all three subsets in the stellar mass-redshift space is shown in Figure \ref{fig:spec}.
\section{Color and Star Formation Evolution}
To characterize the impact of dwarf–dwarf interactions on galaxy evolution, we examined rest-frame colors spanning the ultraviolet, optical, and near-infrared, along with star formation rates. These complementary diagnostics provide insight into recent star formation, stellar population evolution, and dust attenuation, while probing different timescales and physical processes. We describe the physical interpretation of each diagnostic below.

The $NUV-r$ color is a sensitive tracer of recent star formation, combining near-ultraviolet emission dominated by young, massive stars with optical light from the bulk of the stellar population. The NUV-band primarily traces stars with ages $<$100 Myr, making $NUV-r$ highly susceptible to recent or ongoing star formation episodes, with even modest enhancements in star formation producing significant blueward shifts in $NUV-r$ \citep{2007ApJS..173..512S,2007ApJS..173..267S,2007ApJS..173..619K,2013ApJ...767...90K, 2020ApJ...889..188T, 2024MNRAS.531.2223P}.

The rest-frame $U-V$ color traces the recent star formation history by combining sensitivity to young and intermediate-age stars in the U-band with the bulk stellar population traced by the V-band. Stellar population synthesis models show that optical colors respond more gradually to changes in star formation than ultraviolet colors \citep{2003MNRAS.344.1000B, 2013ARA&A..51..393C}. Therefore, while the $NUV-r$ color is more sensitive to very recent activity, $U-V$ reflects star formation averaged over longer timescales of several hundred Myr \citep{2005ApJ...618...91K,2005ApJ...619..193M,2012ApJ...754L..14S,burnham2026itscomplicatedthinkforward}.

The $g-r$ color traces the relative contribution of intermediate-age and older stellar populations in galaxies. The g-band samples the blue optical continuum while the r-band probes redder optical light dominated by longer-lived stars, making $g-r$ primarily sensitive to the mean stellar population age and, to a lesser extent, dust attenuation and metallicity \citep{2001AJ....122.1861S,2003ApJS..149..289B,2012MNRAS.427.2376H,2018MNRAS.475..624N}. Redder $g-r$ colors generally correspond to older or more metal-rich stellar populations, while bluer values indicate younger average stellar ages. Compared to UV-optical colors such as $NUV-r$, which respond strongly to recent star formation, $g-r$ is sensitive to star formation over timescales of a few Gyr. It is therefore better interpreted as an indicator of the overall stellar population age rather than instantaneous star formation activity \citep{2009ARA&A..47..159B}.

Colors that combine optical and near-infrared bands, such as $r-K$ and $V-J$, are sensitive to older stellar populations and particularly sensitive to dust reddening, as dust strongly attenuates optical light while leaving near-infrared emission relatively unaffected. In star-forming systems, reddening in $r-K$ or $V-J$ often indicates increased dust production associated with enhanced star formation rather than aging stellar populations \citep{2007ApJ...655...51W,2015A&A...575A..74S,2009ApJ...691.1879W,2011ApJ...735...53P,2019ApJ...880L...9L,2013A&A...558A..67A,2016A&A...590A.103M,2018MNRAS.479.2147M,2018A&A...617A..70S}. These diagnostics are therefore essential for distinguishing between genuinely quiescent systems and dust-obscured starbursts.

\begin{figure*}[b]
\plotone{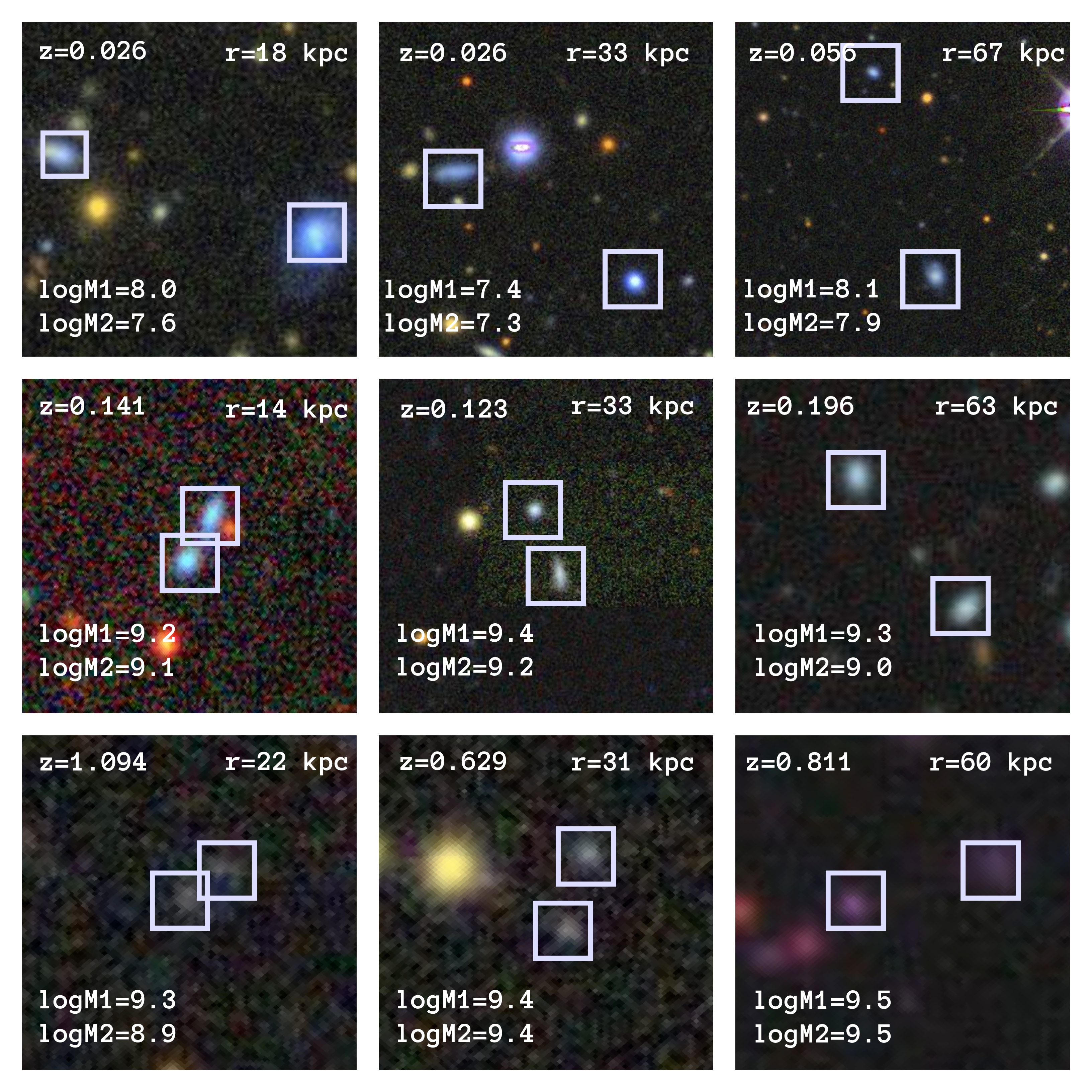}
	\caption{DESI Survey color images of nine dwarf pairs. The top row is low-mass/low-redshift pairs, the middle row is high-mass/low-redshift pairs, and the bottom row is high-mass/high-redshift pairs. Each image shows redshift, projected separation, and stellar masses. \label{figw}}
\end{figure*}

Together, these colors and star formation rate measurements provide a comprehensive view of the evolutionary properties of interacting dwarf galaxies. $NUV-r$ captures rapid, recent star formation; $U-V$ and $g-r$ probe more extended star formation histories; $r-K$ and $V-J$ constrain dust and older stellar populations; and star formation rate directly measures ongoing stellar mass growth. By examining their joint behavior as a function of projected separation, we can trace the timing, intensity, and duration of interaction-driven processes across different mass and redshift regimes. DESI Legacy Survey color images of randomly selected pairs from all three subsets are shown in Figure \ref{figw}.
\subsection{Low-mass/low-redshift}
\begin{figure}[ht!]
\plotone{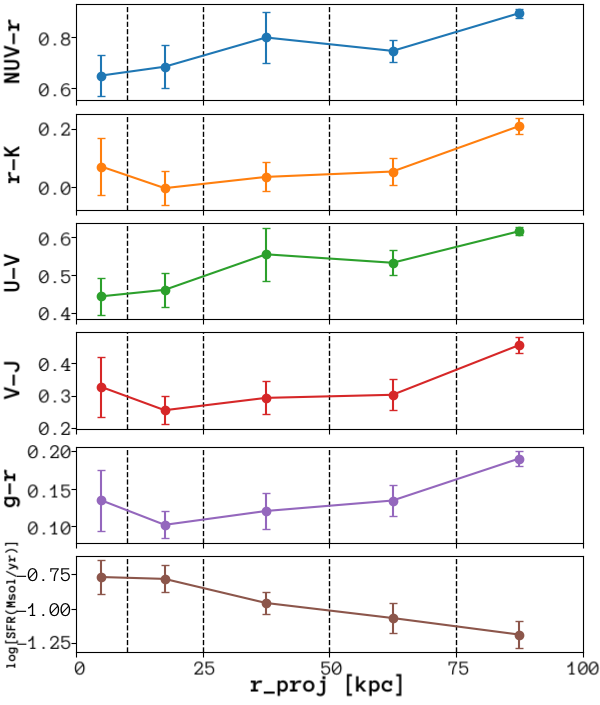}
	\caption{Median color and star formation rate evolution of low-mass/low-redshift dwarfs with respect to projected separation.
	\label{fig4}}
\end{figure}

\begin{figure}[ht!]
\plotone{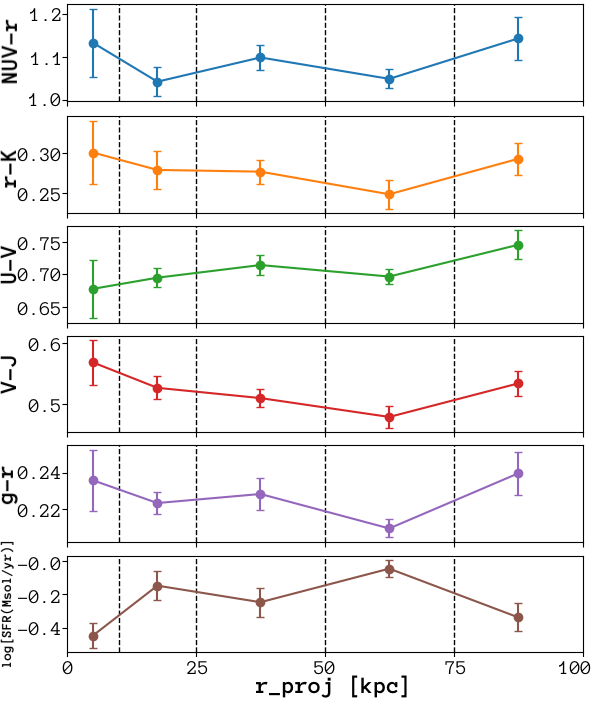}
	\caption{Median color and star formation rate evolution of high-mass/low-redshift dwarfs with respect to projected separation. \label{fig5}}
\end{figure}
\begin{figure}[ht!]
\plotone{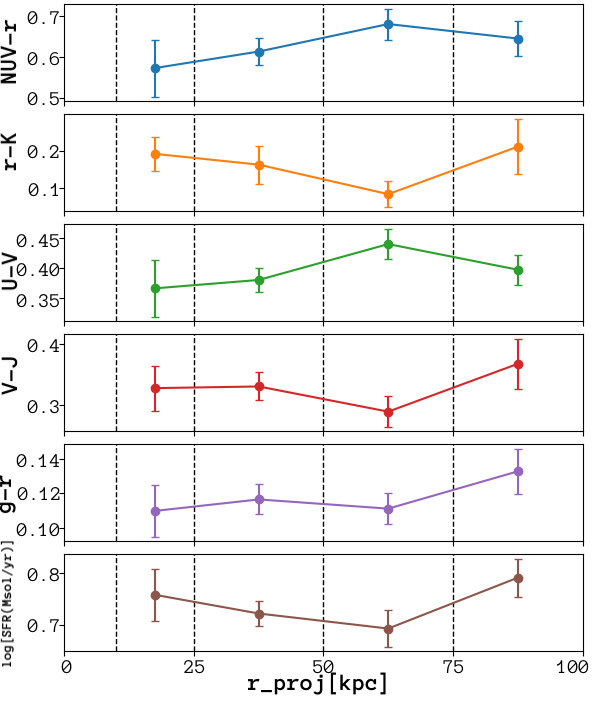}
	\caption{Median color and star formation rate evolution of high-mass/high-redshift dwarfs with respect to projected separation. No pairs are available in the smallest separation bin (0-10 kpc) due to resolution limitations. \label{fig66}}
\end{figure}
The low-mass/low-redshift dwarf pairs exhibit a clear and coherent response to the decreasing projected separation shown in Figure \ref{fig4}. As pair separation decreases from the 75-100 kpc to the 10-25 kpc bin, the star formation rate increases, while all color indices become progressively bluer, indicating interaction-triggered star formation. This behavior is most pronounced at separations of 10-25 kpc, where the differences in star formation rate and colors relative to the 75-100 kpc bin are largest. For low-mass galaxies, this separation is comparable to the median virial radius \citep{2015ApJ...805....2S}, suggesting that tidal interactions are most effective when companions first penetrate each other's halo.

At the smallest separations (0–10 kpc), the trends become more complex and are strongly influenced by uncertainties. Although the star formation rate evidently remains elevated, hints of a trend and potential divergence can be noticed among the colors. The UV–optical colors ($NUV–r$ and $U–V$) show indications of continued blueward evolution, in accordance with elevated star formation rates. In contrast, optical–near-infrared colors ($r–K$ and $V–J$) show evidence of potential reddening in the innermost bin, indicating an increase in dust attenuation. This divergence implies a transition to a phase of dust-obscured star formation, in which newly formed stars are embedded within dusty regions.

The potential reddening observed in the $g–r$ color at the smallest separations is consistent with this scenario. Compact, dusty systems can suppress the g-band emission, while embedded star formation contributes little to the observed optical light. As a result, $g–r$ can redden even as star formation continues, consistent with a centrally concentrated and dust-enshrouded starburst phase.

Taken together, these trends suggest an evolutionary sequence in low-mass dwarf interactions: an initial phase of unobscured, tidally induced star formation at intermediate separations, followed by a compact, dust-obscured starburst at the smallest separations. This behavior highlights the importance of dust-obscured star formation for stellar mass growth even in low-mass systems and suggests that close dwarf–dwarf interactions may contribute to obscured star formation in the local universe, analogous to similar processes inferred at higher redshift \citep{2011A&A...534A..81R,2022ApJ...928...31S,2023MNRAS.518.6142A}.
\subsection{High-mass/low-redshift}
High-mass/low-redshift dwarf pairs exhibit a somewhat different interaction-driven evolutionary sequence compared to their lower-mass counterparts, shown in Figure \ref{fig5}. From the 75–100 kpc bin to the 50–75 kpc bin, the star formation rate increases and all colors become systematically bluer, indicating the onset of tidally triggered star formation. This separation range roughly corresponds to the typical virial radius of galaxies in this mass regime \citep{2015ApJ...805....2S}, analogous to the peak response observed at smaller separations in the low-mass sample.

At intermediate separations (10-25 kpc and 25-50 kpc bins), star formation rates and $NUV-r$, $U-V$, and $g-r$ colors remain relatively flat, suggesting an interaction phase during which gas inflows and star formation proceed without dramatic escalation, while modest reddening in $r-K$ and $V-J$ suggests increased dust attenuation. This behavior is consistent with longer dynamical times and more efficient feedback in higher-mass dwarfs, which can suppress strongly bursty responses.

In the smallest separation bin (0–10 kpc), the trends change abruptly. The star formation rate declines sharply, accompanied by reddening in nearly all broadband colors. Notably, $NUV–r$ becomes redder, indicating quenching of very recent star formation on short timescales. In contrast, $U–V$ does not redden, implying that the intermediate-age stellar population has not yet responded to the shutdown of star formation. This combination suggests a rapid end of star formation rather than a gradual exhaustion. Simultaneous reddening in optical–near-infrared colors indicates increased dust attenuation in the most compact pairs.

Overall, high-mass dwarf interactions appear to follow a sequence in which star formation is first enhanced at halo-entry scales, then regulated, and finally rapidly suppressed at the smallest separations. This contrasts with the low-mass sample, where star formation remains elevated and increasingly obscured at the smallest separations, highlighting a strong mass dependence in the outcomes of dwarf–dwarf interactions.
\subsection{High-mass/high-redshift}
The high-mass/high-redshift subset exhibits an interaction sequence distinct from that of low-redshift pairs, shown in Figure \ref{fig66}. At projected separations of 50–75 kpc, we observe a decrease in the star formation rate, accompanied by reddening in the ultraviolet–optical colors ($NUV–r$, $U–V$). In contrast, the optical–near-infrared colors ($g–r$, $r–K$, $V–J$) become systematically bluer at this separation, a pattern suggestive of reduced internal dust attenuation.

This behavior can be explained by an early stage of the interaction, in which loosely bound gas and dust are removed from the outer regions of galaxies. Such a scenario has observational support from nearby interacting dwarf systems. Observational studies of interacting local volume dwarfs have demonstrated that tidal forces during close encounters can strip loosely bound gas from the outer regions. Although it has been removed from the galaxy, this gas remains bound to the galaxy and can later be reaccreted to fuel episodes of star formation \citep{2016MNRAS.459.1827P}. In gas-rich environments characteristic of high-redshift dwarfs \citep{2013ApJ...776...34K}, tidal and ram-pressure effects may remove diffuse gas and dust before triggering central gas inflows, leading to a transient phase of star-formation suppression and reduced dust content.

At smaller separations, star formation rates rise again and ultraviolet–optical colors become bluer, while optical–NIR colors exhibit reddening. This transition suggests that, after an early stripping event, the displaced material is driven inward by tidal torques, thereby igniting central star formation and increasing dust obscuration. The resulting pattern points to a two-stage interaction pathway in high-mass/high-redshift dwarfs: an initial gas/dust removal phase followed by reaccretion and centrally concentrated star formation and dust buildup.
\section{non-isolated major dwarf interactions}
\begin{figure}[ht!]
\plotone{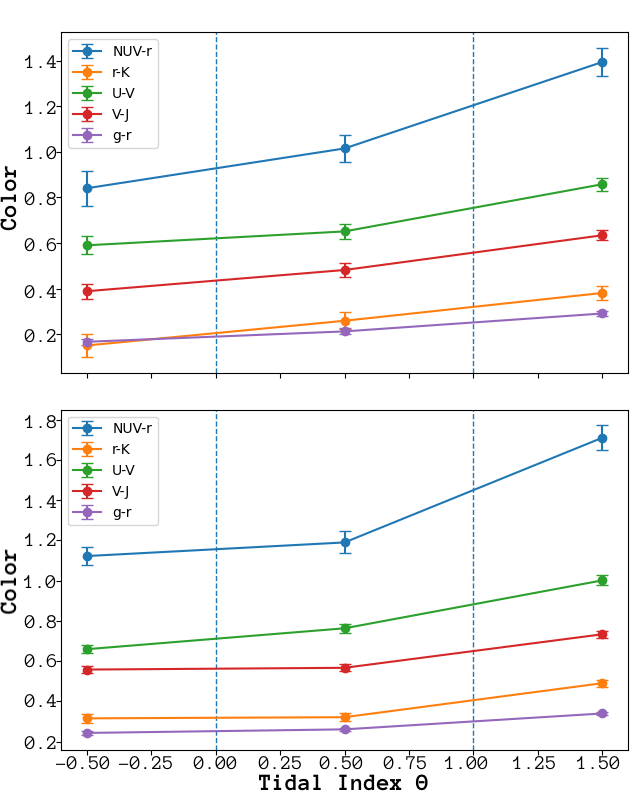}
	\caption{Top: low-mass/low-redshift, bottom: high-mass/low-redshift. Color evolution of non-isolated low-redshift paired dwarfs with respect to tidal index $\Theta$. Increasing $\Theta$ signifies denser environment.\label{fig8}}
\end{figure}
To investigate the impact of the environment on dwarf-dwarf interactions, we constructed a sample of non-isolated dwarf pairs defined as systems with at least one non-dwarf galaxy within a projected distance of 1 Mpc. The non-isolated dwarf pair selection follows the same methodology as the isolated sample, ensuring consistency in the stellar mass, redshift, and projected separation bins. Applying this selection results in 229 low-mass/low-redshift dwarf pairs, 486 high-mass/low-redshift pairs, and 10 high-mass/high-redshift pairs. Because of the small size of the high-redshift subset, the analysis presented here focuses only on the low-redshift regime.

To quantify the strength of the external tidal field, we compute the tidal index $\Theta$, which measures the relative tidal influence of the massive neighbor, and is defined as $\Theta=log \frac{M_* [\times10^8 M_\odot]}{D^3 [Mpc]}$, where M$_*$ and D are the mass and distance of the non-dwarf neighbor \citep{1998A&A...331..891K, 2011ApJ...743....8W}. Thus, a higher tidal index corresponds to a stronger external tidal influence from a more massive and/or closer neighbor, while a lower tidal index indicates a weaker external tidal environment. Across both low-mass and high-mass subsets, we find that galaxy colors systematically redden with increasing tidal index (Figure \ref{fig8}), indicating that a stronger environmental impact is associated with greater dust attenuation and/or reduced star formation \citep{2025MNRAS.538..153K}. This result demonstrates that the tidal index captures physically meaningful environmental effects and justifies its use as a secondary parameter alongside projected separation.

To disentangle the relative roles of dwarf–dwarf interactions and external tidal fields, we examined color and star formation trends as a function of projected separation for two environmental regimes: systems with median $\Theta \approx 0$ and median $\Theta \approx 1$. The choice of the two median $\Theta$ values was motivated by the need to retain sufficient sample sizes for robust statistical analysis in each bin. This approach allows us to isolate how interactions proceed under weak versus strong external tidal influence.
\subsection{Low-mass/low-redshift}
\begin{figure}[ht!]
\plotone{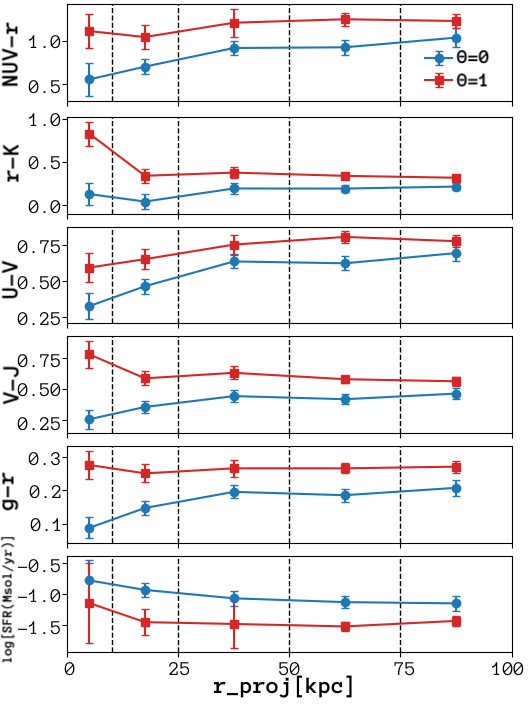}
	\caption{Color and star formation evolution of non-isolated low-mass/low-redshift dwarf pairs with respect to projected separation for two tidal indices: $\Theta=0$ and $\Theta=1$, labeled with blue and red curves, respectively.\label{fig9}}
\end{figure}

For low-mass/low-redshift pairs with median $\Theta \approx 0$, the evolution with decreasing projected separation resembles that of the isolated sample, shown in Figure \ref{fig9}. Star formation rates increase toward smaller separations, accompanied by blueing in $NUV–r$ and $U–V$ colors. These trends are consistent with interaction-triggered star formation. However, in contrast to isolated systems, $V–J$ and $g–r$ also become bluer, indicating negligible dust attenuation and spatially extended star formation. This behavior likely reflects the influence of a weak external tidal field that perturbs the gas, puffs up disks, and prevents efficient gas inflow to the central regions, hence reducing the buildup of compact, dust-enshrouded starbursts.

On the other hand, low-mass/low-redshift pairs with median $\Theta \approx 1$ exhibit a different evolutionary path. From the widest separations down to the 10–25 kpc bin, $NUV-r$ and $U-V$ remain broadly consistent within the uncertainties, with a possible blueward trend, while the other colors remain relatively flat. In the final 0-10 kpc bin, however, the behavior changes, with strong reddening in $r-K$ and $V-J$, accompanied by reddening in $g-r$ and $NUV-r$. Only $U-V$ remains relatively blue. This behavior in the final phase of the interaction resembles that observed in isolated high-mass/low-redshift dwarf pairs, indicating that strong external tides can drive low-mass systems into a dynamical state typically associated with higher-mass systems. The strong reddening in dust-sensitive colors implies heavy dust obscuration, while the reddening of $NUV–r$ suggests the onset of star formation quenching. The persistence of blue $U–V$ colors indicates that this transition occurs on short timescales, such that intermediate-age stellar populations still dominate the optical light.

These results demonstrate that low-mass dwarfs are highly susceptible to environmental tides, with weak tides impacting where star formation happens and how dust is distributed, and strong tides fundamentally altering the mode of interaction-driven evolution.
\subsection{High-mass/low-redshift}
\begin{figure}[ht!]
\plotone{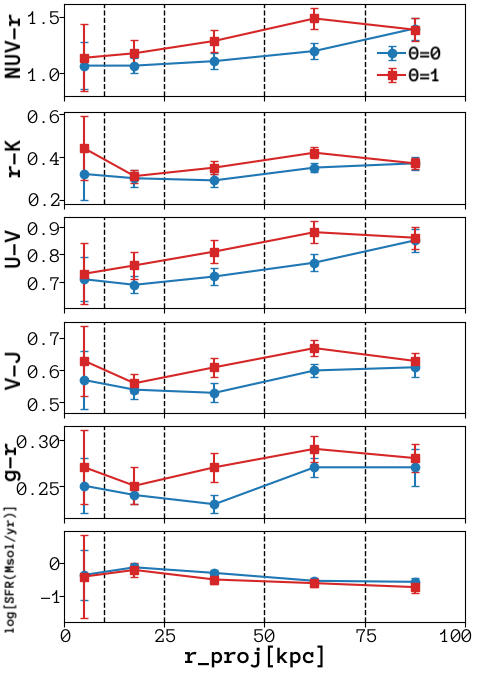}
	\caption{Color and star formation evolution of non-isolated high-mass/low-redshift dwarf pairs with respect to projected separation for two tidal indices: $\Theta=0$ and $\Theta=1$, labeled with blue and red curves, respectively.\label{fig10}}
\end{figure}

We next examine the evolution of colors and star formation rates with decreasing projected separation for high-mass/low-redshift dwarf pairs in both environmental regimes, shown in Figure \ref{fig10}. Over most separation bins, all colors become systematically bluer, and the star formation rate increases mildly, consistent with interaction-driven enhancement of star formation. Despite the similarity in star formation rates between the two subsamples, the colors diverge in the 50-75 kpc bin and later reconverge in the 10-25 kpc bin, with $\Theta=1$ subsample colors being systematically redder during the divergence phase.

At the smallest projected separations, the uncertainties increase substantially, limiting our ability to draw firm conclusions regarding the behavior of recent star formation traced by $NUV–r$ and $U–V$. Although median trends in these colors become less well constrained, dust-sensitive colors such as $r–K$ and $V-J$ show potential reddening in the closest pairs, indicating increased dust attenuation during late interaction stages.

\section{Discussion}
\subsection{Environmental modulation}
These results reveal a strong mass dependence in environmental sensitivity. Low-mass dwarfs are highly susceptible to environmental tides, with the tidal index controlling whether interactions lead to extended star formation, compact dusty starbursts, or rapid suppression of recent star formation. Low-mass pairs exposed to mild tidal influence show interaction-triggered star formation similar to their isolated counterparts, but with no evidence of significant dust buildup. This demonstrates that, unlike in isolated low-mass pairs, where star formation is compact and dust-obscured, mild tides alter the evolution by affecting where star formation occurs and how dust is distributed.

Low-mass pairs under strong tidal impact exhibit evolution in the final stages of the interaction similar to high-mass dwarf pairs. This indicates that strong external tides can drive low-mass systems into a dynamical state typically associated with higher-mass systems. The environment can, therefore, accelerate dwarf galaxy evolution by effectively mimicking the physics of higher-mass interactions, an effect that is likely important for understanding dwarf pre-processing in group environments and at earlier cosmic epochs \citep{2015ApJ...808L..27W}. Additionally, low-mass pairs under strong tidal influence are systematically redder and have lower star-formation rates than their counterparts under mild tidal influence.

Even though high-mass dwarf pairs in both $\Theta=1$ and $\Theta=0$ types of environment exhibit identical star formation rates and symmetric color trends (Figure \ref{fig10}), a significant divergence in colors occurs in the 50-75 kpc bin with $\Theta=1$ subsample being systematically redder during the divergence phase. The presence of color offsets at fixed star formation rate suggests that environmental effects primarily modulate dust attenuation rather than the instantaneous level of star formation. The stronger tidal field in $\Theta=1$ systems likely enhances gas compression and increases dust column density. At smaller separations, dwarf–dwarf tidal interactions dominate the gas dynamics in both subsamples, erasing prior environmental differences and yielding similar dust content and stellar population properties. The reconvergence of colors in the 10-25 kpc bin indicates that internal interaction-driven processes govern the galaxy response in this mass regime. This behavior is consistent with the deeper gravitational potentials and higher gas surface densities of higher-mass dwarfs, which stabilize gas disks against strong redistribution by external tides \citep{2011ApJ...726...98K}.
\subsection{The role of dust}
Across many regimes, we find evidence for increased dust build-up during the final stages of the interaction. The reddening in the $r–K$ and $V–J$ colors, even when the UV colors remain blue, indicates that a significant fraction of the interaction-triggered star formation in dwarfs may be hidden from UV-based diagnostics. This has important implications for understanding the contribution of dwarf galaxies to dust-obscured star formation at intermediate and high redshift, and suggests that dwarf–dwarf mergers may represent a previously understudied pathway for the formation of dust-obscured galaxies in the early universe \citep{2022ApJ...936..118Y}.

Compact, intensely star-forming, dust-reddened cores identified in the 0-10 kpc bin of low-mass interacting dwarfs potentially represent immediate progenitors of blue compact dwarfs (BCD). They are currently in the dusty starburst phase, where feedback processes have not yet cleared the line of sight. As the starburst continues, the galactic winds will eventually remove the obscuring dust, leaving behind a compact, blue core. \cite{2020AJ....159..103K} found that isolated dwarf-dwarf post-mergers, the stage of interaction that follows the 0-10 kpc stage, exhibit properties in line with BCDs. Therefore, the interaction sequence examined in this paper provides a consistent evolutionary picture in which interacting dwarfs are transformed into BCDs.

High-mass/low-redshift pairs also show dust buildup; however, the star formation rates decline sharply in the final bin. Physically, this suggests that although tidal torques efficiently drive gas and dust towards the central regions, feedback processes can halt star formation. This decoupling between star formation rate and dust content implies that dust buildup is governed not only by instantaneous star formation, but also by the redistribution and concentration of the interstellar medium during interactions.
\subsection{Divergent final-stage evolution of low- and high-mass dwarfs}
As previously mentioned, low- and high-mass dwarf pairs exhibit different star formation outcomes in the final stages of the interaction: low-mass systems maintain elevated star formation rates, while high-mass systems undergo abrupt quenching. This divergence suggests a mass-dependent response of star formation to the late stages of dwarf–dwarf interactions, potentially reflecting differences in the interplay between interaction-driven gas inflows, gas consumption, and stellar feedback. In low-mass dwarfs, continued gas inflow may sustain star formation through the final interaction stages. In contrast, the deeper gravitational potentials of high-mass dwarfs may allow the buildup of dense central gas reservoirs, where rapid gas consumption and feedback can subsequently suppress star formation on short timescales. Thefore, the observed divergence may arise from a different balance between gas inflows, consumption, and feedback in the two mass regimes.

A mass dependence in the star formation response to interactions is also seen in hydrodynamical simulations. Using the NewHorizon simulation, \cite{2021MNRAS.500.4937M} found that lower-mass galaxies exhibit higher interaction-induced star formation enhancements than higher-mass galaxies. More generally, FIRE simulations demonstrate that stellar feedback strongly regulates star formation in dwarf galaxies and produces bursty star formation histories and substantial short-timescale variability \citep{2014MNRAS.445..581H}. Feedback-driven gas outflows and inflows can also produce repeated cycles of gas redistribution in low-mass galaxies \citep{2016ApJ...820..131E}. While these simulations do not directly reproduce the specific interaction sequence observed here, they demonstrate that the interplay between gas inflows and feedback can produce strongly time-dependent star formation in dwarf galaxies.
\subsection{Comparison to massive interacting galaxies}
The trends observed in interacting dwarf galaxies broadly resemble those seen in more massive systems, where galaxy interactions drive enhanced star formation, with the enhancement becoming stronger toward smaller projected separations and continuing into post-merger stage \citep{1996ApJ...471..115B,2008AJ....135.1877E,2013MNRAS.435.3627E,2020MNRAS.493.3716H,Shah_2022}. In both regimes, close pairs exhibit increased dust attenuation, with optical–near-infrared colors reddening as star formation becomes progressively obscured \citep{1996ARA&A..34..749S}. The reddening of $r-K$ and $V-J$ colors at small separations in our sample is consistent with this picture, suggesting that even dwarf galaxies enter a dust-obscured star-forming phase during interactions.

However, dwarf-dwarf interactions appear to exhibit a more complex evolutionary response than their massive counterparts. Rather than following a relatively uniform interaction sequence, their evolutionary outcomes depend strongly on stellar mass and are modulated by the external environment. This complexity reflects the greater susceptibility of low-mass systems to dynamical perturbations and feedback-driven regulation owing to their shallower gravitational potentials \citep{2019MNRAS.485.3317S,2022NatAs...6..647C}
\section{Conclusions}
In this work, we presented the largest systematic study to date of dwarf–dwarf interactions across a broad range of stellar masses ($6.5 \leq \log(M_\star/M_\odot) \leq 9.5$), redshifts $0.015 \leq z \leq 1.5$, projected separations ($r_{proj}\leq100$ kpc), and environmental conditions (strong and weak external tidal fields), using the DESI EDR. By identifying over a thousand dwarf pairs, we extend the study of dwarf interactions well beyond the low-redshift, relatively massive dwarf regime that has dominated previous works. Our results demonstrate that the traditional picture of dwarf–dwarf interactions as uniformly star formation–enhancing is overly simplistic.

We find that interaction-driven evolution in dwarf galaxies is sensitive to three coupled parameters: stellar mass, interaction stage, as traced observationally by projected separation, and the strength of the external tidal field. While enhanced star formation and bluer UV colors are common outcomes of interactions, whether star formation is extended versus compact, unobscured versus dust-enshrouded, sustained versus rapidly quenched varies strongly across this parameter space.

Low-mass/low-redshift dwarf pairs, regardless of the environment, exhibit enhanced star formation accompanied by bluing of all colors with reduced projected separation, consistent with interaction-triggered, dust-unobscured starbursts. However, in the final stages of the interaction, when the separation is the smallest, evolution begins to diverge: isolated pairs undergo compact dust-obscured star formation, pairs exposed to mild tidal impact of the environment exhibit spatially extended star formation, while pairs exposed to strong external tidal forces show evidence of heavy dust obscuration and abrupt quenching of star formation, similar to their higher-mass counterparts. This suggests that strong environmental tides can accelerate dwarf evolution by shifting low-mass systems into an interaction regime more typical of higher-mass galaxies.

Similarly, high-mass/low-redshift dwarf pairs in all three environmental regimes show evidence of interaction-triggered star formation and overall bluer colors across the spectrum. In the final stages of the interaction, isolated pairs exhibit increased dust obscuration and rapid quenching. Simultaneously, in pairs exposed to mild and strong external impact, increased dust presence can be noticed, while it is unclear what happens to star formation due to large uncertainties. Therefore, in this regime, separation-dependent trends are largely insensitive to the strength of the external tidal field, indicating that, above a critical mass threshold, the internal dynamics of the interaction dominate the response.

Finally, high-mass/high-redshift pairs exhibit behavior consistent with loosely bound gas being displaced from the galaxies in the early stages of the interaction, and later reaccreted and used as fuel for subsequent star-forming episodes. These dynamics has been observed and described in detail in a few local volume dwarf-dwarf pairs.

Our results show that dwarf–dwarf interactions are not governed by a single evolutionary pathway, in agreement with some recent studies of individual systems \citep[e.g.,][]{2024ApJ...963...37K}. Instead, they span a range of outcomes shaped by mass, environment, and the interaction stage. Given that dwarf galaxies dominate the galaxy population at early cosmic times, and the merger rate per unit volume, the complex behaviors uncovered here likely represent fundamental building blocks of early galaxy evolution. 

\begin{acknowledgments}
MM thanks Xinyu Dai and Erin Kado-Fong for insightful discussions.
\end{acknowledgments}





\bibliography{dids}{}
\bibliographystyle{aasjournalv7}

\end{document}